# Hofstadter Butterfly and Broken-Symmetry Quantum Hall States in $\alpha$-Type Organic Dirac Fermion Systems

Toshihito Osada*

*Institute for Solid State Physics, University of Tokyo, Kashiwa, Chiba 277-8581, Japan.*

The electronic state under magnetic fields in the $\alpha$-type organic Dirac fermion systems, $\alpha$-(ET)$_2$I$_3$ and $\alpha$-(BETS)$_2$I$_3$, has been studied to clarify the spatial order in the quantum Hall state. The four-band tight-binding model with Peierls phase factors was employed, and the generated Hofstadter butterfly and its Chern numbers confirmed the validity of the Dirac fermion picture in these materials. The four-component envelope function of the $N = 0$ Landau level with valley degeneracy was investigated. It was found that the two degenerate valley states have different weights on A and A' molecules connected by inversion. This valley-site correspondence is also recognized for the $N = 0$ spin-split Landau levels under the Zeeman effect and the spin-orbit interaction. The spontaneous valley symmetry breaking in the $N = 0$ Landau levels due to the exchange interaction results in the $\nu = \pm 1$ quantum Hall states accompanied by the spatial modulation of charge and spin densities at A and A' sites in a unit cell.

## 1. Introduction: Broken-Symmetry Quantum Hall States

Layered organic conductors $\alpha$-(BEDT-TTF)$_2$I$_3$ and $\alpha$-(BEDT-TSeF)$_2$I$_3$, which are often abbreviated as $\alpha$-(ET)$_2$I$_3$ and $\alpha$-(BETS)$_2$I$_3$, are considered to be two-dimensional (2D) Dirac fermion (DF) systems under high pressure [1-3]. On their conducting layers, ET or BETS molecules form an anisotropic triangular lattice called the $\alpha$-type configuration where the unit cell contains four molecules. Four 2D $\pi$-bands are constructed from the HOMO of these molecules. Under appropriate pressure, the third and fourth bands have no overlap and touch at two points, $\mathbf{k}_0$ and $-\mathbf{k}_0$, in the Brillouin zone forming two tilted type-I Dirac cones (valleys). The Fermi level is stoichiometrically located at the Dirac point without any carrier doping. The emergence of 2D DF state with type-I Dirac cones under pressure has been experimentally confirmed for both real systems by the $\pi$-Berry phase of Shubnikov-de Haas oscillation using doped samples with contact charge [4, 5].

At ambient pressure, $\alpha$-(ET)$_2$I$_3$ undergoes a metal-insulator transition at 135K due to charge ordering caused by electron correlation. On the other hand, $\alpha$-(BETS)$_2$I$_3$ undergoes a metal-insulator crossover at around 50K, which is believed to be a result of the small topological insulator gap due to relatively strong spin-orbit interaction (SOI) [3, 6]. Appropriate high pressure can suppress the insulating behavior of both systems, resulting in the metallic 2D DF state. Although their band parameters differ slightly, the qualitative feature of their DF state is expected to be similar, except for the effect of SOI.

Recently, the $\nu$ =1 quantum Hall (QH) plateau was observed in the Hall resistance of $\alpha$-(BETS)$_2$I$_3$ under high pressure [7]. This is the first observation of the QH effect in a bulk single crystal of organic conductors, which is caused by the finite natural electron doping. Moreover, it is the $\nu$ =1 QH effect of a 2D DF system with type-I Dirac cones having $\pi$-Berry phase. In general, the 2D DF system in which the Landau levels (LLs) have four-fold spin and valley degeneracy shows the QH effect with Hall plateau $\sigma_{xy} = -\nu(e^2/h)$ with $\nu = \pm2, \pm6, \pm10, ...$ . Although the Zeeman effect breaks the spin degeneracy, the $\nu = 1$ QH effect is expected only when both spin and valley degeneracy are broken in the $N = 0$ LL. The valley degeneracy could be broken in the bulk crystal by the electron-electron interaction, more precisely by the exchange interaction. In the spin-split $N = 0$ LL with valley degeneracy, the Pauli principle prevents two electrons from occupying the same valley at a given center coordinate. Therefore, the repulsive interaction between two electrons belonging to the same valley is effectively weakened. This leads to the spontaneous breaking of the valley degeneracy. One valley is selectively occupied, leading to filling-dependent valley splitting. The $\nu = 1$ QH effect is expected when the $N = 0$ LL with one spin is about half-filled and spontaneously splits into two valley levels with the Fermi level in between.

In the conventional 2D electron systems with multiple valleys, such as the $n$-channel inversion layer of Si MOSFET, the QH effect due to valley splitting has also been observed [8]. However, their valley splitting does not originate from the spontaneous degeneracy breaking, but from the inter-valley coupling under the confinement potential, although the splitting is enhanced by the exchange interaction [9].

In graphene, which is the typical 2D massless DF system on the honeycomb lattice, the broken-symmetry QH states have been studied experimentally and theoretically [10-16]. Graphene's LLs are believed to have a four-fold spin and valley degeneracy due to the small Zeeman splitting compared to the LL spacing. Several possible states have been theoretically discussed, particularly for the $\nu = 0$ QH states, which result from the degeneracy breaking of the $N = 0$ LL. These include the spin-polarized (ferromagnetic) states, the canted antiferromagnetic state (CAF), the bond order (intervalley coherent) state with Kekule distortion (BO), the charge density wave state (CDW) [11], and the CAF-BO coexistence state [14, 15]. Additionally, the STM technique has visually observed the spatial modulation pattern of charge density in these broken-symmetry states [16]. Graphene has two carbon sites (A and B) in a unit cell, and two $\pi$-bands touching at two Brillouin zone corners (K and K' points), forming two Dirac cones (valleys). The $N = 0$ LL at the K(K') valley consists only of the wave function at the B(A) site. This valley-site correspondence causes the spatial charge and/or spin modulation in a unit cell in broken-symmetry QH states.

The similar spatial modulation is also expected in the $\nu = 1$ QH state in the $\alpha$-type organic DF system with a different lattice structure and relatively large Zeeman splitting [17]. The key to this problem is the presence and aspect of the valley-site correspondence in the $N = 0$ LL of the $\alpha$-type organic conductor.

**2. Tight-Binding Model with Peierls Phase and Spin-Orbit Interaction**

In this paper, we discuss the real-space charge/spin order pattern in the broken-symmetry QH states of the $\alpha$-type organic DF system. Unlike graphene, each conducting layer of the $\alpha$-type organic conductors has four molecular sites (A, A', B, and C) in a unit cell on the anisotropic triangular lattice, as shown in Fig. 3 (b). To clarify the spatial structure of the wave functions under magnetic fields, we use the 2D four-band tight-binding model for the $\alpha$-type organic conductors [1, 2] instead of the two-band **k**-linear effective model, i.e., the tilted Weyl model [18, 19]. Additionally, we take into account the SOI, which is considered significant in $\alpha$-(BETS)$_2$I$_3$ [3].

In a 2D electron system with in-plane potential modulation, a moving electron experiences an effective magnetic field perpendicular to the plane. Therefore, the SOI is dependent only on the normal spin component $s_z$ (Ising-type SOI) [20]. Thus, the Hamiltonian of the system is decoupled into two spin sectors as $H(\mathbf{k}) = H(\mathbf{k}, s_z = +1) \oplus H(\mathbf{k}, s_z = -1)$. In the $\alpha$-type organic DF system, $H(\mathbf{k}, s_z)$ is a $4 \times 4$ matrix including the dimensionless SOI strength $\lambda$ at zero magnetic field [21]. The bases of the matrix are the Bloch sums constructed from HOMOs of molecules A, A', B, and C.

Under the normal magnetic field $\mathbf{B} = (0, 0, B)$, transfer integrals in the Hamiltonian obtain the Peierls phase factors. These factors extend the original unit cell to the magnetic unit cell. We choose the Landau gauge $\mathbf{A} = (0, Bx, 0)$ and consider the magnetic field that satisfies the following relation.

$$\frac{\Phi_{uc}}{\Phi_0} = \frac{eBab}{2\pi\hbar} = \frac{p}{q} = \frac{4p'}{q} \ . \tag{1}$$

Here, $p'$ and $q$ are coprime integers and $p = 4p'$. $\Phi_{uc}$ represents the magnetic flux passing through a unit cell, and $\Phi_0 = 2\pi\hbar/e$ is the flux quantum. $b$ and $a$ are the lattice constants in the $x$- and $y$-direction, respectively. The factor 4 on the right side reflects the fact that the crystal unit cell (eight triangular plaquettes) is four times larger than the unit cell of the molecular triangular lattice (two plaquettes), which determines the Peierls phase factors. The magnetic unit cell ($S_{muc} = qb \times a$) consists of $q$ original unit cells ($S_{uc} = b \times a$) denoted by $j$ (= 1, ..., $q$) aligned in the $x$-direction. In the magnetic field, the bases of the tight-binding Hamiltonian $H(\mathbf{k}, s_z)$ are the Bloch sum constructed from four types of molecules in a magnetic unit cell. Thus $H(\mathbf{k}, s_z)$ becomes the following $4q \times 4q$ matrix.

$$H(\mathbf{k}, s_z) = \begin{pmatrix} M^{(1)} & N^{(1,2)} & & & & {}^tN^{(0,1)}* \\ {}^tN^{(1,2)}* & M^{(2)} & N^{(2,3)} & & & \\ & {}^tN^{(2,3)}* & M^{(3)} & N^{(3,4)} & & \\ & & {}^tN^{(3,4)}* & M^{(4)} & \ddots & \\ & & & \ddots & \ddots & N^{(q-1,q)} \\ N_{0,1} & & & & {}^tN^{(q-1,q)}* & M^{(q)} \end{pmatrix}. \quad (2)$$

Here, the diagonal blocks are given by the following $4 \times 4$ matrices.

$$M^{(j)} = \begin{pmatrix} \varepsilon_0 + \Delta + s_z\mu_B B & M_{AA'}^{(j)}(\mathbf{k}) & M_{AB}^{(j)}(\mathbf{k}, s_z) & M_{AC}^{(j)}(\mathbf{k}, s_z) \\ M_{AA'}^{(j)}(\mathbf{k})^* & \varepsilon_0 - \Delta + s_z\mu_B B & M_{A'B}^{(j)}(\mathbf{k}, s_z) & M_{A'C}^{(j)}(\mathbf{k}, s_z) \\ M_{AB}^{(j)}(\mathbf{k}, s_z)^* & M_{A'B}^{(j)}(\mathbf{k}, s_z)^* & \varepsilon_0 + s_z\mu_B B & M_{BC}^{(j)}(\mathbf{k}) \\ M_{AC}^{(j)}(\mathbf{k}, s_z)^* & M_{A'C}^{(j)}(\mathbf{k}, s_z)^* & M_{BC}^{(j)}(\mathbf{k})^* & \varepsilon_0 + s_z\mu_B B \end{pmatrix}, \quad (3)$$

$$M_{AA'}^{(j)}(\mathbf{k}) = a_2 e^{+i\pi\frac{\Phi_{uc}}{\Phi_0}j} e^{+i\frac{ak_y}{2}} + a_3 e^{-i\pi\frac{\Phi_{uc}}{\Phi_0}j} e^{-i\frac{ak_y}{2}}$$

$$M_{BC}^{(j)}(\mathbf{k}) = a_1 e^{+i\pi\frac{\Phi_{uc}}{\Phi_0}\left(j+\frac{1}{2}\right)} e^{+i\frac{ak_y}{2}} + a_1 e^{-i\pi\frac{\Phi_{uc}}{\Phi_0}\left(j+\frac{1}{2}\right)} e^{-i\frac{ak_y}{2}}$$

$$M_{AB}^{(j)}(\mathbf{k}, s_z) = b_2 e^{-i\frac{\pi}{2}\frac{\Phi_{uc}}{\Phi_0}\left(j+\frac{1}{4}\right)} (1+i\lambda s_z) e^{i\left(+\frac{bk_x}{2} - \frac{ak_y}{4}\right)}$$

$$M_{AC}^{(j)}(\mathbf{k}, s_z) = b_1 e^{+i\frac{\pi}{2}\frac{\Phi_{uc}}{\Phi_0}\left(j+\frac{1}{4}\right)} (1+i\lambda s_z) e^{i\left(+\frac{bk_x}{2} + \frac{ak_y}{4}\right)} \quad .$$

$$M_{A'B}^{(j)}(\mathbf{k}, s_z) = b_3 e^{+i\frac{\pi}{2}\frac{\Phi_{uc}}{\Phi_0}\left(j+\frac{1}{4}\right)} (1-i\lambda s_z) e^{i\left(+\frac{bk_x}{2} + \frac{ak_y}{4}\right)}$$

$$M_{A'C}^{(j)}(\mathbf{k}, s_z) = b_4 e^{-i\frac{\pi}{2}\frac{\Phi_{uc}}{\Phi_0}\left(j+\frac{1}{4}\right)} (1-i\lambda s_z) e^{i\left(+\frac{bk_x}{2} - \frac{ak_y}{4}\right)}$$

As for the transfer integrals of the $\alpha$-type organic conductor shown in Fig. 3(b), we use $a_1 = -0.038$ eV, $a_2 = +0.080$ eV, $a_3 = -0.018$ eV, $b_1 = +0.123$ eV, $b_2 = +0.146$ eV, $b_3 = -0.070$ eV, and $b_4 = -0.025$ eV, which have been often used for the DF state in $\alpha$-(ET)$_2$I$_3$ under uniaxial pressure $P_a = 0.4$ GPa [1]. In the diagonal elements of $M^{(j)}$, A and A' site potentials ($\pm\Delta$) are added to introduce valley splitting by inversion symmetry breaking. The Zeeman energy $s_z\mu_B B$ is also added, where $\mu_B$ is the Bohr magneton and Lande's g-factor is assumed to be $g = 2$. The SOI parameter $\lambda$ is introduced to preserve time reversal symmetry at zero field according to Ref. [21]. The off-diagonal blocks of $H(\mathbf{k}, s_z)$ are given by the following $4 \times 4$ matrices.

$$N^{(j-1,j)} = \begin{pmatrix} 0 & 0 & 0 & 0 \\ 0 & 0 & 0 & 0 \\ N_{BA}^{(j-1,j)}(\mathbf{k},s_z) & N_{BA'}^{(j-1,j)}(\mathbf{k},s_z) & 0 & 0 \\ N_{CA}^{(j-1,j)}(\mathbf{k},s_z) & N_{CA'}^{(j-1,j)}(\mathbf{k},s_z) & 0 & 0 \end{pmatrix}, \quad (4)$$

$$N_{BA}^{(j-1,j)}(\mathbf{k},s_z) = b_3 e^{+i\frac{\pi}{2}\frac{\Phi_{uc}}{\Phi_0}\left(j-\frac{1}{4}\right)}(1+i\lambda s_z)e^{i\left(+\frac{bk_x}{2}+\frac{ak_y}{4}\right)}$$

$$N_{BA'}^{(j-1,j)}(\mathbf{k},s_z) = b_2 e^{-i\frac{\pi}{2}\frac{\Phi_{uc}}{\Phi_0}\left(j-\frac{1}{4}\right)}(1-i\lambda s_z)e^{i\left(+\frac{bk_x}{2}-\frac{ak_y}{4}\right)}$$

$$N_{CA}^{(j-1,j)}(\mathbf{k},s_z) = b_4 e^{-i\frac{\pi}{2}\frac{\Phi_{uc}}{\Phi_0}\left(j-\frac{1}{4}\right)}(1+i\lambda s_z)e^{i\left(+\frac{bk_x}{2}-\frac{ak_y}{4}\right)}$$

$$N_{CA'}^{(j-1,j)}(\mathbf{k},s_z) = b_1 e^{+i\frac{\pi}{2}\frac{\Phi_{uc}}{\Phi_0}\left(j-\frac{1}{4}\right)}(1-i\lambda s_z)e^{i\left(+\frac{bk_x}{2}+\frac{ak_y}{4}\right)}$$

By diagonalizing $H(\mathbf{k},s_z)$, energy levels under the magnetic field $\Phi_{uc}/\Phi_0 = 4p'/q$, which we call the Hofstadter levels, are obtained for a wave number $\mathbf{k}$ and a spin $s_z$. Here, $\mathbf{k}$ is a wave number in the magnetic Brillouin zone ($|k_x| \leq \pi/qb$, $|k_y| \leq \pi/a$), and $k_y$ is related to the center coordinate $X_0$ by $X_0 = -l^2 k_y$, where $l = \sqrt{\hbar/eB}$ is the magnetic length. $X_0$ indicates the location of the real-space orbital motion. Hofstadter levels lose $\mathbf{k}$-dispersion at large $q$ where the magnetic Brillouin zone becomes much smaller than the original Brillouin zone. Therefore, we calculate the Hofstadter levels only for $\mathbf{k} = 0$ by choosing a sufficiently large value of $q$ [22]. In addition, we assume infinitesimal $\Delta$ to differentiate valley states by introducing small valley splitting.

### 3. Validity of Dirac Fermion Picture in $\alpha$-Type Organic Conductors

First, we consider only the orbital effect, i.e., the spinless case in which the SOI and Zeeman effect are ignored. Figure 1(a) shows the band dispersion at zero magnetic field, and Fig. 1(b) shows the calculated magnetic field dependence of energy levels, so-called "Hofstadter butterfly" spectrum [23], of the $\alpha$-type organic conductor. The horizontal axis indicates the normalized magnetic field $\Phi_{uc}/\Phi_0 = eBab/2\pi\hbar$.

Although only the fundamental period is plotted in the figure, the energy spectra vary with the period of $\Delta(\Phi_{uc}/\Phi_0) = 8$, which is about 33000 T in real $\alpha$-type conductors. The black points indicate Hofstadter levels, and the other regions correspond to energy gaps, whose color indicates the Chern number without spin degeneracy. Kishigi and Hasegawa first obtained the Hofstadter butterfly spectrum of the $\alpha$-type organic conductor while studying the pressure dependence of the electronic state, in particular, the critical state with the "three-quarter" Dirac point [24, 25]. In the present work, we investigate the valley-site correspondence aspect in the $N = 0$ LL using the site-dependent wave functions obtained by the similar Hofstadter calculation, and the effect of SOI on it.

The Chern number $N_{Ch}$, which is the topological number that gives the QH conductivity $\sigma_{xy} = -N_{Ch}(e^2/h)$, was calculated according to the original TKNN theory [26]. In the magnetic field $\Phi_{uc}/\Phi_0 = p/q$ ($p = 4p'$), the electron density of $r$ occupied Hofstadter levels $n = r/S_{muc}$ can be written as the linear combination of the electron densities of one band and one LL as $n = s(1/S_{uc}) + t(1/2\pi l^2)$ using integer coefficients $s$ and $t$, which leads to the Diophantine equation $sq + tp = r$. According to the Streda formula [27], the Hall conductivity is given by $\sigma_{xy} = (-e)(\partial n/\partial B)_\mu = -t(e^2/h)$. Therefore, we can obtain the Chern number of each gap as $N_{Ch} = t$ by solving $sq + tp = r$ under $|t| \leq q/2$.

We focus on the vicinity of the contact points between the third band $E_3(\mathbf{k})$ and the fourth band $E_4(\mathbf{k})$ in the low magnetic fields, which is indicated by the red circle in Fig. 1(b). The detailed spectra are shown in Fig. 2 on an enlarged scale. As shown in Fig. 2(a), there exist two valleys in the energy region between the van Hove singularities of $E_3(\mathbf{k})$ and $E_4(\mathbf{k})$, which are distinguished by the contact points $-\mathbf{k}_0$ and $+\mathbf{k}_0$. In this region, the energy is quantized into LLs under magnetic fields. Each LL consists of $2 \times p$ Hofstadter levels, where the factor 2 corresponds to the valley degeneracy. In Fig. 2(b) we can see that the LL at the contact point has no field independence (zero mode) and other LLs show a square-root-like field dependence. Moreover, the Chern numbers between neighboring LLs take values of $N_{\text{Ch}} = \pm 1, \pm 3, \pm 5, \cdots$. These features are characteristic of the 2D massless DF system in the spinless case. Importantly, they are obtained directly from the four-band tight-binding model without any approximation using the **k**-linear Weyl model. Therefore, it has been confirmed that the $\alpha$-type conductor behaves like the massless DF system between the van Hove singularities even when the multiband effect is included. The similar confirmation has also been made for graphene [22]. Outside the van Hove singularities, each LL shows double splitting, reflecting the disappearance of the valley degeneracy, as seen in Fig. 2(b).

**4. Valley-Site Correspondence in the *N* = 0 Landau Level**

To investigate the possible real-space pattern of electron density at the broken-symmetry QH state, we analyze the LL wave function at each molecular site for each valley. The wave function of each LL state is obtained by the above diagonalization of $H(\mathbf{k}, 0)$. The values of the envelope function at A, A', B, and C sites in the magnetic unit cell are obtained from the eigenvector for each Hofstadter level. For sufficiently low magnetic fields, one LL for each valley is formed from $1/(2\pi l^2) = p/S_{\text{muc}}$ states with different center coordinate $X_0 = -l^2 k_y$, corresponding to $p \,(= 4p')$ Hofstadter levels with different $(k_x, k_y)$. Thus, for a fixed $X_0$ and a single valley, the probability density at each site is obtained as the sum of the square norm of the envelope functions of the degenerated $p$ Hofstadter levels.

Figure 3(a) shows the probability density of the site-dependent envelope functions of LLs between the van Hove singularities at a low magnetic field $\Phi_{uc}/\Phi_0 = 4/499$ (approximately 33 T). The baseline height of the probability density represents the LL energy, which is denoted by $N = 0, \pm 1, \pm 2, \cdots$. The band dispersion at zero magnetic field is also shown as a function of $X - X_0 = -l^2 k_y$. We can see that the envelope functions correspond to the semiclassical cyclotron orbits surrounding two valleys in **k**-space. This reflects the two-fold valley degeneracy of the LLs. Note the asymmetry of the envelope functions of A and A' sites between $-\mathbf{k}_0$- and $+\mathbf{k}_0$-valleys. In particular, in the $N = 0$ LL at the Dirac point energy, the probability density of A (A') site is larger in the $+\mathbf{k}_0$-valley ($-\mathbf{k}_0$-valley) than in the $-\mathbf{k}_0$-valley ($+\mathbf{k}_0$-valley). Namely, in the $N = 0$ LL, the $+\mathbf{k}_0$-valley is rich in A, while the $-\mathbf{k}_0$-valley is rich in A'. This is nothing but the valley-site correspondence in the $\alpha$-type organic DF system, although it is not perfect, unlike graphene. On the other hand, the probability density of B and C sites is symmetric between $-\mathbf{k}_0$ and $+\mathbf{k}_0$-valleys.

In $\alpha$-type systems, A and A' sites are connected by the inversion operation and are equivalent under inversion symmetry. However, once the inversion symmetry is broken, the $N = 0$ LL splits into the $-\mathbf{k}_0$ and $+\mathbf{k}_0$-valley levels because A and A' sites are no longer symmetric. If the valley-degenerate $N = 0$ LL is occupied almost half, the degeneracy could be spontaneously broken due to exchange interaction. In the spinless case, this generates the broken-symmetry $\nu = 0$ QH state where either A or A' site is charge-rich in a unit cell.

## 5. Effect of Spin-Orbit Interaction

Next, we will discuss the spinful case in which we take into consideration account the SOI and Zeeman effect. In $\alpha$-type organic conductors, the Zeeman energy $s_z \mu_B B$ cannot be ignored relative to the LL spacing because the group velocity is much smaller than in graphene. Therefore, within the one-body effect, $\alpha$-type organic conductors exhibit a significant spin splitting caused by the Zeeman effect, which is further modified by the SOI. However, we will ignore the Zeeman effect when discussing the wave functions to clarify the contribution of the SOI, because the Zeeman effect only shifts the electron energy by $s_z \mu_B B$ without changing the wave functions.

Figure 4(a) shows the Hofstadter butterfly for the $\alpha$-type system with finite SOI ($\lambda > 0$), excluding the Zeeman shift. The red and blue patterns correspond to the spectra of $H(\mathbf{k}, +1)$ and $H(\mathbf{k}, -1)$, respectively. At zero magnetic field, a gap opens due to the SOI, and the system becomes the quantum spin Hall (QSH) insulator, which is a 2D topological insulator with spin conservation [21]. The SOI gap has Chern numbers $N_{\text{Ch}}^{(s_z=+1)} = +1$ and $N_{\text{Ch}}^{(s_z=-1)} = -1$ for $s_z = +1$ and $s_z = -1$, respectively, as confirmed by comparing with Fig. 1 at finite low fields. Therefore, the zero-field QSH insulator state continuously changes to the spin-polarized $\nu = 0$ QH state (QH ferromagnetic state) under finite magnetic fields. The similar feature has also been studied in graphene [28].

Figure 4(b) shows the probability density of site-dependent envelope functions at a low magnetic field $\Phi_{\text{uc}} / \Phi_0 = 4/499$ (approximately 33 T) for the system with finite SOI, without considering the Zeeman splitting. The $N = 0$ LL causes clear spin splitting into the "$N = 0 \uparrow$" ($s_z = -1$) and "$N = 0 \downarrow$" ($s_z = +1$) levels, located at the bottom of $E_4(\mathbf{k})$ and the top of $E_3(\mathbf{k})$, respectively. Note that this correspondence is reversed by changing the sign of the SOI parameter $\lambda$. The $N = 0 \uparrow$ and $N = 0 \downarrow$ LLs still have two-fold valley degeneracy. Other LLs also show small spin splitting, with valley degeneracy. The asymmetry of envelope functions of A and A' sites between $-\mathbf{k}_0$- and $+\mathbf{k}_0$-valleys (valley-site correspondence) is observed especially in the $N = 0 \uparrow$ and $N = 0 \downarrow$ LLs, even under finite spin splitting. Therefore, the SOI does not affect the valley-site correspondence, although it modifies the spin splitting due to the Zeeman effect.

## 6. Spatial Order of Charge and Spin Density in the $\nu = \pm 1$ Quantum Hall States

Thus far, the Zeeman effect has not been taken into account. However, even if we consider a finite Zeeman effect, the structure of the wavefunction remains unchanged. Only the value and sign of the spin splitting change. Therefore, the above argument still holds, except for the value of the spin splitting.

When the "$N = 0 \uparrow$" or "$N = 0 \downarrow$" LL is almost half-occupied, the exchange interaction spontaneously breaks the valley degeneracy and creates a filling-dependent energy gap, resulting in the emergence of the $\nu = 1$ or $\nu = -1$ QH state as shown in Fig. 4(d). These QH states are accompanied by spatial charge and spin modulation in a unit cell, breaking the inversion symmetry. Figure 4(c) schematically illustrates this feature when spin splitting is mainly due to the SOI. Therefore, we can conclude that the asymmetry of charge and spin densities on A and A' molecules spontaneously appears in the $\nu = \pm 1$ broken-symmetry QH states of the $\alpha$-type organic conductors.

This spatial modulation may be detected experimentally using STM or NMR. In graphene, the STM measurement visually observed the real-space modulation pattern of the broken-symmetry QH states, including the BO and CDW states [16]. If the DF state can be realized in the $\alpha$-type organic conductor in a vacuum, the spatial modulation of charge density can be observed by STM in the QH states. The use of $^{13}$C-NMR in the QH states may be a more effective method for detecting spatial spin modulation because four molecular sites give different NMR peaks. At zero magnetic field, in fact, previous site-resolved NMR measurements have successfully detected the ferrimagnetic spin polarization in $\alpha$-(ET)$_2$I$_3$ [29].

## 7. Conclusions

In summary, we studied the broken-symmetry QH states in $\alpha$-type organic 2D DF systems using the four-band tight-binding model instead of the two-band linearized model. The Hofstadter butterfly and its Chern numbers confirmed the conventional DF picture. We investigated the four-component envelope function of spin-split $N = 0$ LLs with valley degeneracy, considering SOI. We found that the two valley states have different weights on A and A' molecules connected by inversion. The exchange interaction spontaneously breaks the valley degeneracy, leading to a spatially modulated $\nu = \pm 1$ QH state with different charge and spin densities on A and A' molecules.

**Acknowledgements**

This work was partially supported by JSPS KAKENHI Grant Numbers JP21K18594 and JP23K03297. The author thanks T. Sekine for helpful comments on the NMR measurement.

*corresponding author, osada@issp.u-tokyo.ac.jp

**Figure 1** (Osada)

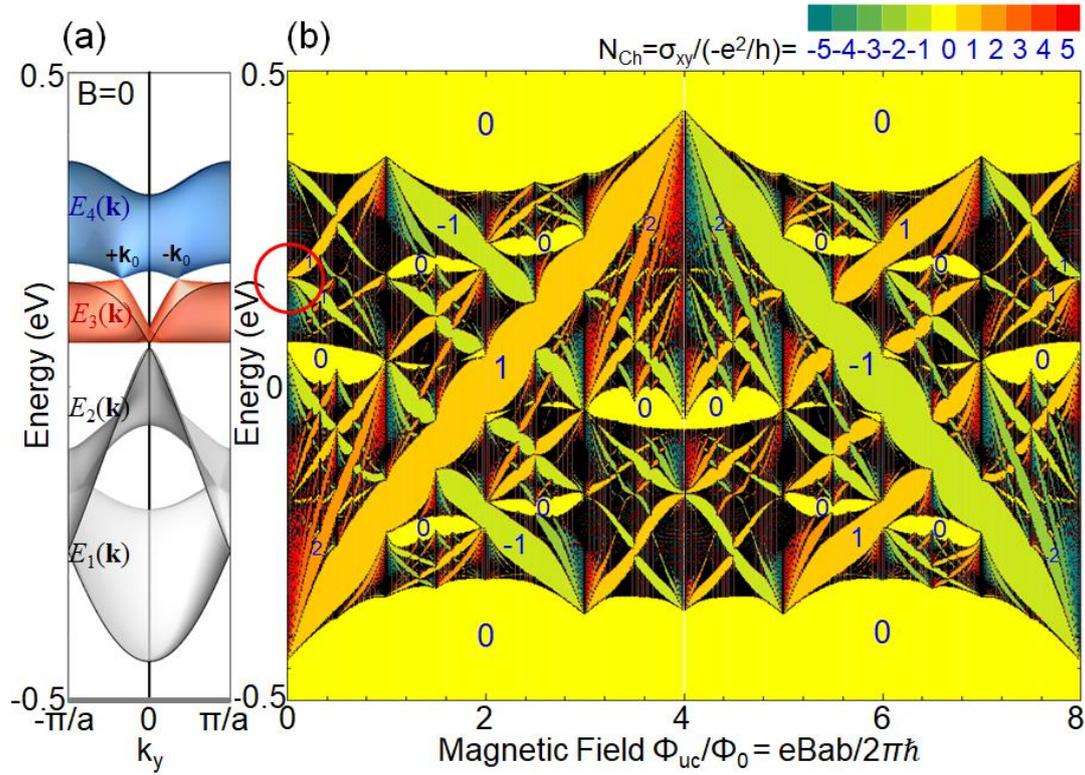

**Figure 1.** (Color online)

(a) Band dispersion of an $\alpha$-type organic conductor $\alpha$-(ET)$_2$I$_3$ under pressure at zero magnetic field. (b) Magnetic field dependence of energy levels (Hofstadter butterfly) in an $\alpha$-type organic conductor $\alpha$-(ET)$_2$I$_3$. The spin degree of freedom is not considered (spinless case). The calculation was done for $q = 251$. The energy gap between quantized levels is colored according to the Chern number $N_{Ch}$.

**Figure 2** (Osada)

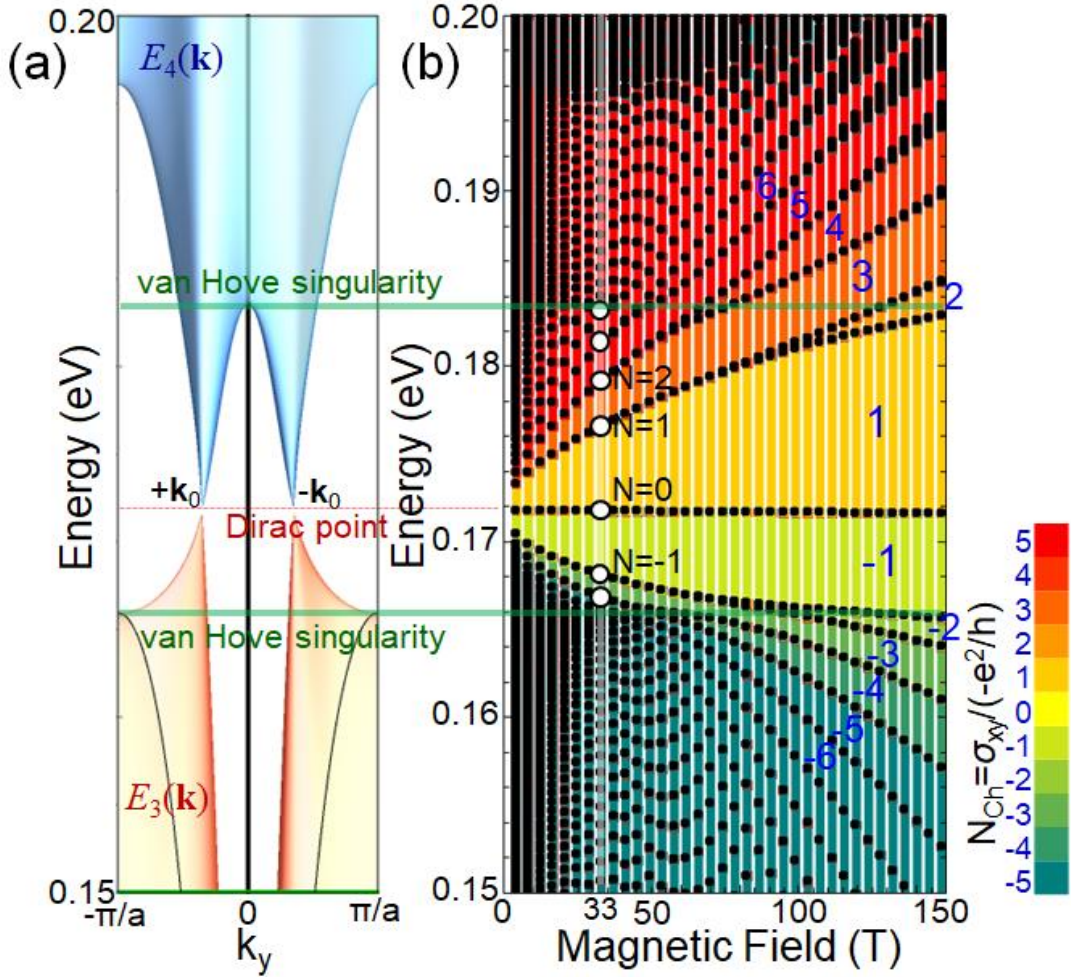

**Figure 2.** (Color online)

(a) Band dispersion around contact points of the third and fourth bands in an $\alpha$-type conductor $\alpha$-(ET)$_2$I$_3$. (b) Magnetic field dependence of energy levels around contact points in the low magnetic field region. The spin degree of freedom is not considered (spinless case). The calculation was done for $q = 4001$. The energy gap between quantized levels is colored by the Chern number $N_{Ch}$.

**Figure 3** (Osada)

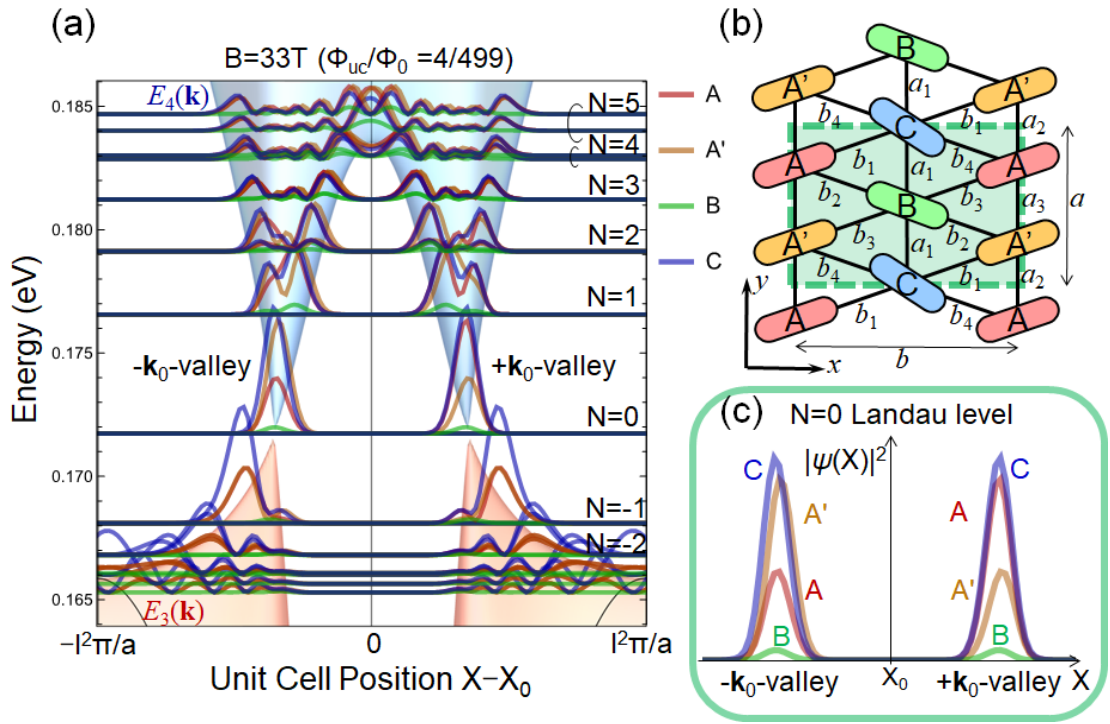

**Figure 3.** (Color online)

(a) Probability densities of the LLs on four molecular sites in the real space at a low magnetic field (approximately 33 T). The spin degree of freedom is not considered. Additionally, the zero-field dispersion is displayed as a function of $X - X_0 = -l^2 k_y$. (b) Schematic crystal lattice of $\alpha$-type organic conductors with transfer integrals. The shaded rectangle indicates a unit cell. (c) Site-resolved probability density of the $N = 0$ LL with two-fold valley degeneracy.

**Figure 4** (Osada)

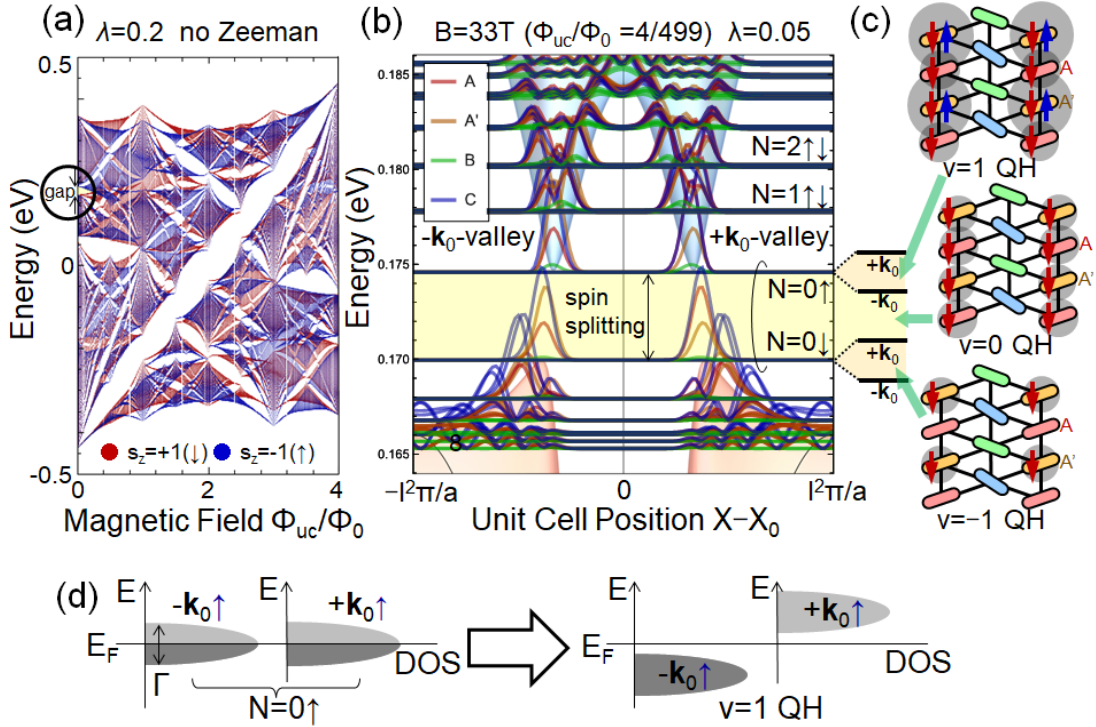

**Figure 4.** (Color online)

(a) Hofstadter butterfly for the $\alpha$-type organic conductor with finite SOI. Zeeman effect is not considered. The color (red and blue) indicates the spin. (b) Probability densities of LLs under the SOI at four molecular sites at a low magnetic field (approximately 33 T). Zeeman effect is not considered. The zero-field gapped dispersion is also shown. (c) Schematics of spontaneous valley splitting of the "$N = 0 \uparrow$" and "$N = 0 \downarrow$" LLs and the examples of the charge and spin density patterns of the broken-symmetry $\nu = \pm 1$ QH states when the SOI is dominant for the spin splitting. (d) Schematic of spontaneous valley splitting of the half-occupied $N = 0 \uparrow$ LL due to exchange interaction. The $\nu = 1$ QH state emerges. $\Gamma$ is the LL width due to scattering.